\documentstyle[preprint,tighten,aps,epsf,12pt,fleqn]{revtex}
\begin{document} 
\preprint{MKPH-T-96-30}
\title{Chiral dynamics of rare eta decays and virtual Compton scattering 
off the nucleon}
\author{S.\ Scherer}
\address{Institut f\"ur Kernphysik, Johannes Gutenberg-Universit\"at,
J.\ J.\ Becher-Weg 45, D-55099 Mainz, Germany}
\date{December 2, 1996}
\maketitle
\begin{abstract}
\thispagestyle{empty}   
   Two different topics relevant for DA$\Phi$NE and CEBAF, respectively,
are covered.
   First we analyze the rare radiative eta decay mode 
$\eta\to\pi\pi\gamma\gamma$ within the framework of chiral perturbation 
theory at ${\cal O}(p^4)$.
   We then discuss virtual Compton scattering off the nucleon at low energies.
   Predictions for the two spin-independent generalized polarizabilities
are shown. 
\end{abstract}
\section{Introduction}
   In the limit of massless $u$, $d$, and $s$ quarks, the QCD Lagrangian
exhibits a global $\mbox{SU(3)}_L\times\mbox{SU(3)}_R$ symmetry which is
assumed to be spontaneously broken to $\mbox{SU(3)}_V$, thus giving
rise to eight pseudoscalar Goldstone bosons which transform under 
$\mbox{SU(3)}_V$ as an octet.
   The lightest pseudoscalar mesons ($\pi,K,\eta$) are commonly identified
with these Goldstone bosons; their finite masses result from an explicit 
chiral symmetry breaking in the QCD Lagrangian due to the finite masses of the 
light quarks.
   Chiral perturbation theory (ChPT) [1,2] provides a systematic framework for 
describing the interactions of these effective degrees of freedom at low 
energies.
   It is based on a description in terms of the most general, effective
Lagrangian consistent with the symmetries of the underlying theory, 
namely, QCD.
   In the meson sector, the effective Lagrangian of ChPT is organized as
a sum of terms with an increasing number of covariant derivatives and
quark mass terms,
\begin{equation}
\label{lmeson}
{\cal L}_{\mbox{\footnotesize eff}}={\cal L}_2+{\cal L}_4 
+ {\cal L}_6 +\cdots,
\end{equation}
where the subscripts refer to the order in the momentum expansion.
   A systematic analysis of Feynman diagrams evaluated with the Lagrangian
of Eq.\ (\ref{lmeson}) is made possible by Weinberg's power counting
scheme [1].
  The lowest-order Lagrangian ${\cal L}_2$ corresponds to a nonlinear
$\sigma$ model with a coupling to external sources. 
   Besides the quark masses, it contains two parameters, namely, the pion 
decay constant in the chiral limit, $F_0\approx 93\,\mbox{MeV}$, and a 
constant $B_0$ related to the quark condensate \mbox{$<\!\bar{q}q\!>$}.
   The Lagrangian ${\cal L}_4$ was determined and studied by Gasser
and Leutwyler [2]. 
   In the $\mbox{SU(3)}$ sector it involves 10 low-energy constants,
some of which have infinite pieces required to compensate infinities 
resulting from one-loop diagrams with vertices of ${\cal L}_2$.
   In addition, at ${\cal O}(p^4)$ one has the Wess-Zumino-Witten action [3]
which accounts for the chiral anomaly of QCD but does not introduce a new 
parameter.
   For strong and electromagnetic processes involving an odd number of
Goldstone bosons, the Wess-Zumino-Witten term provides the leading-order
contribution.
   At ${\cal O}(p^6)$ the effective Lagrangian involves 111 terms of
even and 32 terms of odd intrinsic parity [4].
   
   Ever since the sixties the interactions of Goldstone bosons with 
baryons have been described in terms of tree-level diagrams derived from 
effective chiral Lagrangians.    
   A consistent power counting, and thus a systematic treatment of 
higher-order corrections has become possible within the framework of the 
heavy-baryon formulation of chiral perturbation theory.
   For example, the pion-nucleon interaction in the one-nucleon sector can be 
described in terms of the most general effective Lagrangian [5]
\begin{equation}
\label{mesonnucleon}
{\cal L}_{\mbox{\footnotesize eff}}=\widehat{\cal L}_{\pi N}^{(1)}
+\widehat{\cal L}_{\pi N}^{(2)} +\widehat{\cal L}_{\pi N}^{(3)} +\cdots,
\end{equation}
   where the individual pieces contain 1, 7, and 24 low-energy constants,
respectively. 
   In particular, the constants of the ${\cal O}(p)$ and ${\cal O}(p^2)$
Lagrangians are scale-independent, i.e., they are not needed to absorb
infinities of loop diagrams. 
   For a review concerning the status of ChPT in the one-nucleon sector
the reader is referred to [6].
   
\section{The rare decay $\eta\to\pi\pi\gamma\gamma$ in chiral perturbation
theory}
   According to [7], the $\eta$ width is $\Gamma = (1.18\pm 0.11)\,\mbox{keV}$,
and the main decay modes are given by $2\gamma$ 
($(39.25 \pm 0.31)$\%), $3\pi^0$ ($(32.1\pm 0.4)$\%), 
$\pi^+\pi^-\pi^0$ ($(23.2\pm 0.5)$\%), and $\pi^+\pi^-\gamma$ ($(4.78\pm
0.12)$\%), where the branching fractions $\Gamma_i/\Gamma$ are given in
parentheses.
   In particular, all other decays appear with a fraction of less than
a percent. 
   With an anticipated number of $\sim 3.2\times 10^8$ etas per year
at DA$\Phi$NE, it will be possible to investigate some of the rare
decays. 
   For example, for the mode $\eta\to\pi^+\pi^-\gamma\gamma$, so far there
exists only an upper limit 
$\Gamma(\eta\to\pi^+\pi^-\gamma\gamma)/\Gamma <2.1\times 10^{-3}$ [7],
whereas for the corresponding neutral mode no information is available.

   We have performed an exploratory investigation of the rare decay modes
$\eta\to\pi\pi\gamma\gamma$ within the framework of chiral perturbation
theory [8].
   At lowest order in the momentum expansion, the invariant amplitude has
been calculated using the Lagrangian
\begin{eqnarray}
\label{leppgg}
{\cal L}_{\mbox{eff}}&=&{\cal L}_2+{\cal L}_{WZW}\nonumber\\
&=&\frac{F^2_0}{4}\mbox{Tr}(D_\mu U (D^\mu U)^\dagger) 
+\frac{F^2_0}{4}\mbox{Tr}(\chi U^\dagger + U\chi^\dagger)\nonumber\\
&&+\frac{e}{16\pi^2}\epsilon^{\mu\nu\alpha\beta}A_\mu
\mbox{Tr}(Q\partial_\nu U\partial_\alpha U^\dagger \partial_\beta U 
U^\dagger-Q\partial_\nu U^\dagger \partial_\alpha U\partial_\beta
U^\dagger U)\nonumber\\
&&-\frac{ie^2}{8\pi^2}\epsilon^{\mu\nu\alpha\beta}\partial_\mu
A_\nu A_\alpha\mbox{Tr}[Q^2(U\partial_\beta U^\dagger
+\partial_\beta U^\dagger U)
-\frac{1}{2}QU^\dagger Q \partial_\beta U +\frac{1}{2}Q U Q \partial_\beta 
U^\dagger],
\end{eqnarray}
where $U(x)=\exp\left(i\phi(x)/F_0\right)$ contains the relevant Goldstone
bosons as well as the singlet eta. 
   In Eq.\ (\ref{leppgg}) we have included only those terms which actually
contribute at ${\cal O}(p^4)$.
   In particular, we have not included ${\cal L}_4$ of Gasser and Leutwyler,
since in an electromagnetic process involving an odd number of Goldstone
bosons it will only contribute at ${\cal O}(p^6)$.
\begin{figure}[h]
\centerline{\epsfxsize=5.6cm\epsfbox{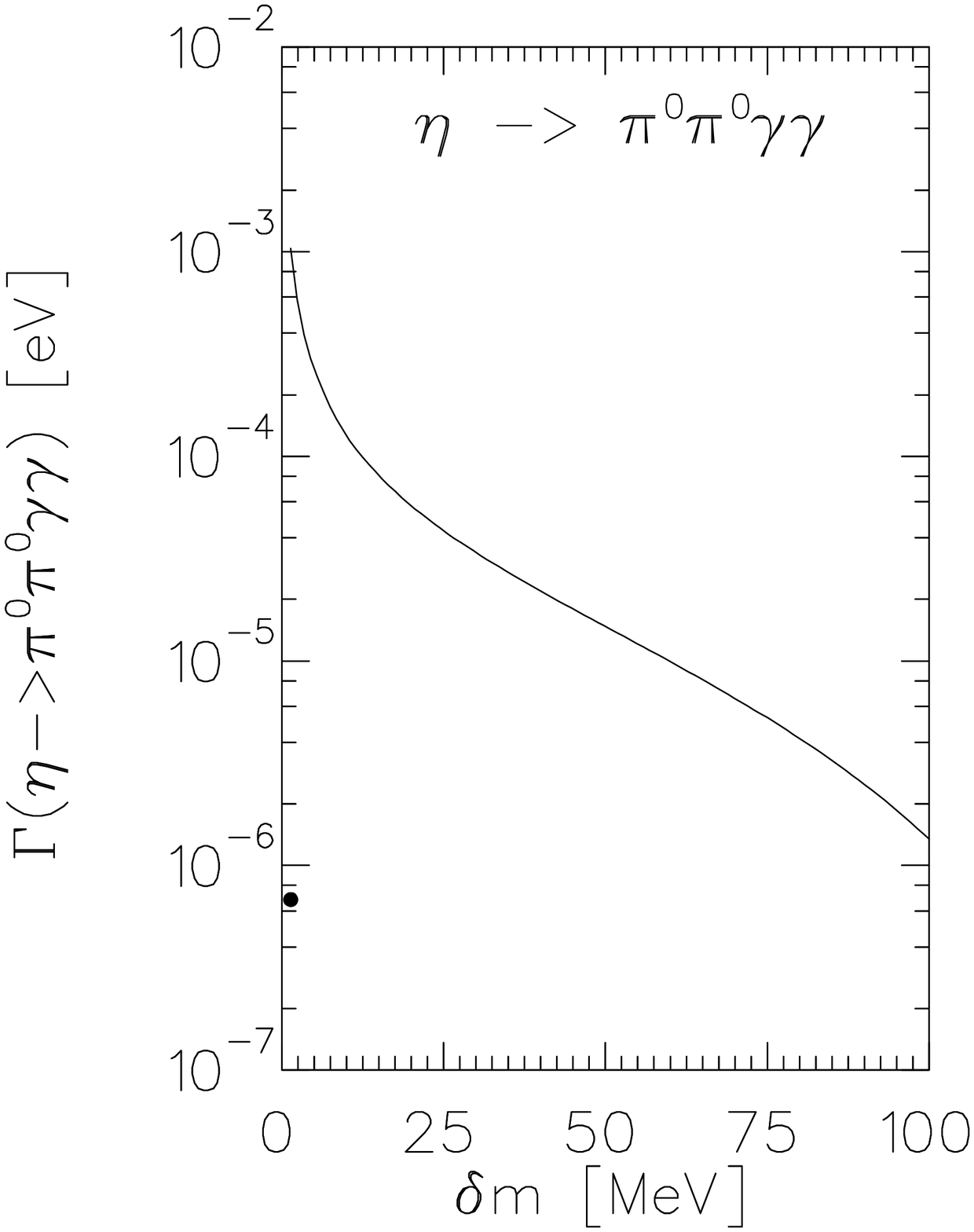}\hspace{1cm}
\epsfxsize=6.0cm\epsfbox{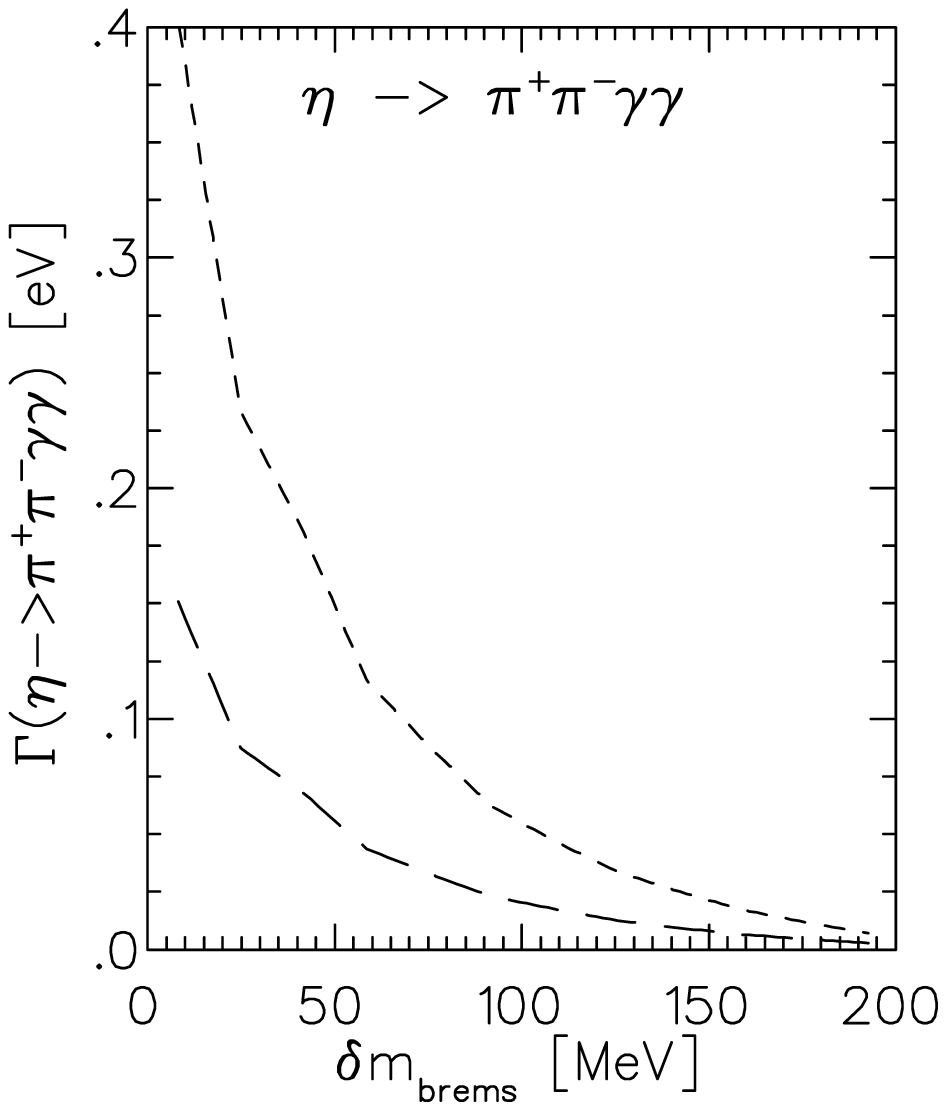}}
\caption{Left figure: Partial decay rate 
$\Gamma(\eta\to\pi^0\pi^0\gamma\gamma)$ as a function of an energy cut
$\delta m$ in $s_\gamma^{1/2}$ around $m_\pi$. The solid line corresponds
to the lowest-order prediction of chiral perturbation theory with $m_u=5$ MeV
and $m_d=9$ MeV. In particular, the decay rate is completely dominated by
the $\pi^0$-pole contribution. The dot corresponds to the isospin-symmetric
case $m_u=m_d=7$ MeV, i.e., without $\pi^0$ pole. 
Right figure: Partial decay rate $\Gamma(\eta\to\pi^+\pi^-\gamma\gamma)$
as a function of an energy cut 
$\delta m_{brems}$ around the bremsstrahlung
singularity at $s_\gamma=0$. The long-dashed line corresponds to the
lowest-order calculation. The dashed line includes vector mesons as 
dynamical gauge boson fields of a hidden local symmetry [10].}
\end{figure}
   Furthermore, terms of the WZW action which are irrelevant for our
purposes have been omitted.
   For the decay constants we use $F_\pi=93\,\mbox{MeV}$, $F_8=1.25\, 
F_\pi$, and $F_1=1.06\, F_\pi$.
   The covariant derivative $D_\mu U =\partial_\mu U +ieA_\mu [Q,U]$ in
${\cal L}_2$ contains a coupling to the electromagnetic field, where 
$Q=\mbox{diag}(\frac{2}{3},-\frac{1}{3},-\frac{1}{3})$ is the quark-charge 
matrix.
   The quark masses are contained in $\chi=2 B_0 M$ with  
$M=\mbox{diag}(m_u,m_d,m_s)$. 
   Finally, $\eta-\eta'$ mixing is taken into account by using a 
phenomenological mixing angle $\theta=-20^\circ$:
\begin{equation}
|\eta>=\cos(\theta)|\eta_8>-\sin(\theta)|\eta_1>,\quad
|\eta'>=\sin(\theta)|\eta_8>+\cos(\theta)|\eta_1>.
\end{equation}

   In Fig.\ 1 (left) the partial decay rate $\Gamma(\eta\to\pi^0\pi^0\gamma
\gamma)$ is shown as a function of an energy cut $\delta m$ in 
$s_\gamma^{1/2}$ around $s_\gamma^{1/2}=m_{\pi^0}$, where $s_\gamma$ is the 
invariant energy squared of the di-photon system.    
   At lowest order, the decay rate is completely dominated by the
$\pi^0$ pole contribution. 
   Using a more realistic strength for the $\eta\pi^0\pi^0\pi^0$ vertex
than predicted by ${\cal L}_2$ leads to an enhancement of the decay
rate $\Gamma(\eta\to\pi^0\pi^0\gamma\gamma)$ of about a factor four.
   Furthermore, as was shown in [9], the inclusion of chiral pion loops
at ${\cal O}(p^6)$ leads to a notable enhancement of the decay rate.
   In particular, in the di-photon spectrum for large values of $s_\gamma$,
these contributions even dominate the $\pi^0$-pole
contribution.
   The $\eta$-pole contribution on the other hand is negligible.
   In Fig.\ 1 (right) the partial decay rate $\Gamma(\eta\to\pi^+\pi^+\gamma
\gamma)$ is shown as a function of an energy cut $\delta 
m_{\mbox{\footnotesize brems}}$ 
around the bremsstrahlung singularity at $s_\gamma =0$.
   The long-dashed line represents the ${\cal O}(p^4)$ calculation,
whereas the dashed line results from a calculation including vector mesons
[10]. 
   The strong enhancement indicates that ${\cal O}(p^6)$ counter terms
may be important.
   The role of loops still has to be investigated.

\section{Virtual Compton scattering off the nucleon at low energies}
   An investigation of the virtual Compton scattering amplitude which,
e.g, enters the physical process $e^-p \to e^-p \gamma$, has recently 
attracted a lot of interest.
   Even though the experiments will be considerably more complicated 
than for real Compton scattering, there is the prospect of obtaining 
completely new information about the structure of the nucleon
which cannot be obtained from any other experiment.
   In particular, as a prerequisite for studying this new structure 
information it is important to identify the model-independent
properties of the VCS amplitude.

   Omitting the Bethe-Heitler contribution, the invariant amplitude for VCS 
reads
\begin{equation}
\label{mvcs}
{\cal M}_{VCS}=-ie^2\epsilon_\mu \epsilon_\nu'^\ast M^{\mu\nu}
=-ie^2\epsilon_\mu M^\mu
=ie^2\left(\vec{\epsilon}_T\cdot\vec{M}_T
+\frac{q^2}{\omega^2}\epsilon_z M_z\right),
\end{equation}
where $\epsilon_\mu=e\bar{u}\gamma_\mu u/q^2$ is the polarization vector
of the virtual photon, and where we made use of current conservation.
   In the center-of-mass system, using the Coulomb gauge for the final
real photon, the transverse and longitudinal parts of ${\cal M}_{VCS}$ 
can be expressed in terms of eight and four independent structures, 
respectively,
\begin{equation}
\vec{\epsilon}_T\cdot \vec{M}_T=\vec{\epsilon}\,'^\ast \cdot 
\vec{\epsilon}_T A_1 + \cdots,\quad
M_z=\vec{\epsilon}\,'^\ast \cdot \hat{q} A_9 + \cdots,
\end{equation}
   where the functions $A_i$ depend on the three kinematical variables
$|\vec{q}|$, $|\vec{q}\,'|$, and $z=\hat{q}\cdot\hat{q}\,'$.

   Model-independent predictions for the functions $A_i$, based on
Lorentz invariance, gauge invariance, crossing symmetry, and the discrete 
symmetries were obtained in [11].
   For example, the result for $A_1$ up to second order in the momenta
$|\vec{q}|$ and $|\vec{q}\,'|$ reads 
\begin{eqnarray}
\label{a1}
A_1&=&-\frac{1}{M}+\frac{z}{M^2}|\vec{q}|
-\left(\frac{1}{8M^3}+\frac{r^2_E}{6M}-\frac{\kappa}{4M^3}
-\frac{4\pi\alpha_0}{e^2}\right)|\vec{q}\,'|^2\nonumber\\
&&+\left(\frac{1}{8M^3}+\frac{r^2_E}{6M}-\frac{z^2}{M^3}
+\frac{(1+\kappa)\kappa}{4M^3}\right)|\vec{q}|^2.
\end{eqnarray}
   To this order, all funtions $A_i$ can be expressed in terms of known 
quantities, namely, $M$, $\kappa$, $G_E$, $G_M$, $r^2_E$, $\alpha_0$, and
$\beta_0$.
   A systematic analysis of the structure-dependent terms specific to VCS 
was first performed in [12].
   The non-pole terms were expressed in terms of three spin-independent
and seven spin-dependent generalized polarizabilities which were introduced
in the framework of a multipole analysis, restricted to 
first order in $|\vec{q}\,'|$, but for arbitrary $|\vec{q}|$.

   A calculation of the spin-independent polarizabilities in the framework
of the linear $\sigma$ model [13] led to the surprising result that only
two of the three polarizabilities were, in fact, independent.
   That this is not a model-dependent statement could be shown in
[14], where the reduction from three to two polarizabilities was traced 
back to a combination of invariance with respect to charge conjugation
and nucleon crossing.
   A similar reduction from seven to four terms occurs for the 
spin-dependent polarizabilities.

   The generalizations of the spin-independent polarizabilities 
$\alpha(|\vec{q}|)$ and $\beta(|\vec{q}|)$ have been calculated in
various frameworks [12,13,15,16] and are shown in Fig.\ 2.
   In particular, chiral perturbation theory at ${\cal O}(p^3)$ 
predicts [16]
\begin{equation}
\label{alphabeta}
\alpha(|\vec{q}|)=\alpha_0\left[1-\frac{7}{50}\frac{|\vec{q}|^2}{m^2_\pi}
+O\left(\frac{|\vec{q}|^4}{m^4_\pi}\right)\right],\quad
\beta(|\vec{q}|)=\beta_0\left[1+\frac{1}{5}\frac{|\vec{q}|^2}{m^2_\pi}
+O\left(\frac{|\vec{q}|^4}{m^4_\pi}\right)\right],
\end{equation}
where $\alpha_0=12.8\times 10^{-4}\,\mbox{fm}^3$ and 
$\beta_0=1.3\times 10^{-4}\,\mbox{fm}^3$ are the results for the real
Compton scattering polarizabilities at ${\cal O}(p^3)$ [6] which compare
remarkably well with the experimental numbers 
$\alpha_0=(12.1\pm 0.8\pm 0.5)\times 10^{-4}\,\mbox{fm}^3$
and $\beta_0=(2.1\pm 0.8 \pm0.5)\times 10^{-4}\,\mbox{fm}^3$ [7].

\begin{figure}[h]
\centerline{\epsfxsize=10cm\epsfbox{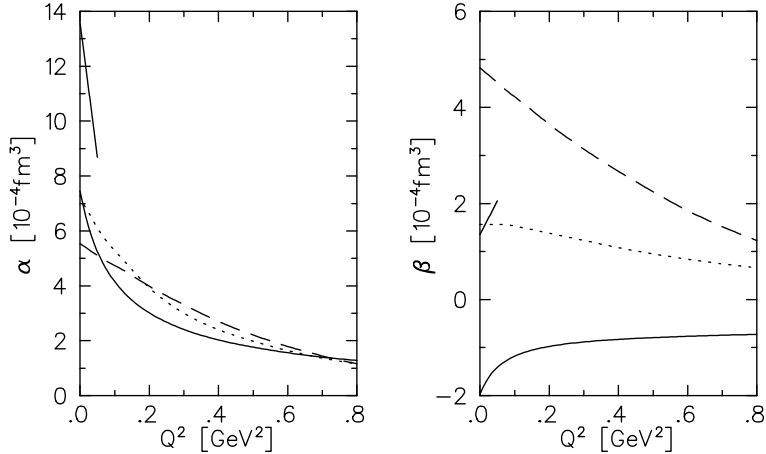}}
\caption{Predictions of different models for $\alpha(Q^2)$ and $\beta(Q^2)$:
linear $\sigma$ model (full line), constituent quark model (dashed line),
effective Lagrangian model (dotted line). The predictions of ChPT are
given by the short lines ending at $Q^2=0.05\,\mbox{GeV}^2$ 
(taken from [13]).}
\end{figure}

\frenchspacing
\noindent {\bf REFERENCES}\\
$[1]$ S. Weinberg, Physica {\bf 96A}, 327 (1979).\\
$[2]$ J. Gasser and H. Leutwyler, Ann. Phys. (N.Y.) {\bf 158}, 142 (1984);
      Nucl. Phys. {\bf B250}, 465 (1995).\\
$[3]$ J. Wess and B. Zumino, Phys. Lett. {\bf 37B}, 95 (1971);
      E. Witten, Nucl. Phys. {\bf B223}, 422 (1983).\\
$[4]$ H. W. Fearing and S. Scherer, Phys. Rev. {\bf D53}, 315 (1996).\\
$[5]$ G. Ecker and M. Moj\v{z}i\v{s}, Phys. Lett. {\bf B365}, 312 (1996).\\
$[6]$ V. Bernard, N. Kaiser, U.-G. Mei{\ss}ner, Int. J. Mod. Phys. {\bf E4}, 193
(1995).\\
$[7]$ R. M. Barnett {\em et al.} (Particle Data Group), Phys. Rev.
{\bf D54}, 1 (1996).\\
$[8]$ G. Kn\"ochlein, S. Scherer, D. Drechsel, Phys. Rev. {\bf D53}, 3634 
(1996).\\
$[9]$ S. Bellucci and G. Isidori, {\tt hep-ph/9610328}.\\
$[10]$ G. Kn\"ochlein and B. R. Holstein (unpublished).\\
$[11]$ S. Scherer, A. Yu. Korchin, J. H. Koch, Phys. Rev. {\bf C54}, 904 
(1996).\\
$[12]$ P. A. M. Guichon, G. Q. Liu, A. W. Thomas, Nucl. Phys. {\bf A591},
606 (1995).\\
$[13]$ A.\ Metz and D.\ Drechsel, MKPH-T-96-08, to appear in Z.\ Phys.\ 
{\bf A}.\\
$[14]$ D. Drechsel, G. Kn\"ochlein, A. Metz, S. Scherer, {\tt nucl-th/9608061},
to appear in Phys. Rev. {\bf C}.\\
$[15]$ M.\ Vanderhaeghen, Phys.\ Lett.\ {\bf B368}, 13 (1996).\\
$[16]$ Th. R. Hemmert, B. R. Holstein, G. Kn\"ochlein, S. Scherer,  
{\tt nucl-th/9608042}, to appear in Phys. Rev. {\bf D}.
\end{document}